\documentclass[conference]{IEEEtran}
\IEEEoverridecommandlockouts
\usepackage{cite}
\usepackage{amsmath,amssymb,amsfonts}
\usepackage{algorithmic}
\usepackage{graphicx}
\usepackage{epstopdf}
\usepackage{textcomp}
\usepackage{xcolor}
\usepackage{diagbox}
\usepackage{array}
\usepackage{amsmath}

\def\BibTeX{{\rm B\kern-.05em{\sc i\kern-.025em b}\kern-.08em
    T\kern-.1667em\lower.7ex\hbox{E}\kern-.125emX}}
\begin{document}

\title{A Quasi-deterministic Channel Model for Underwater Acoustic Communication Systems
}

\author{{Yuxuan Yang\textsuperscript{1},  Yilin Ma\textsuperscript{1,2 }, Hengtai Chang\textsuperscript{2,1 *}, Cheng-Xiang Wang\textsuperscript{1,2 *}}
	\\
	\textsuperscript{1}National Mobile Communications Research Laboratory, School of Information Science and Engineering, \\
	Southeast University, Nanjing 210096, China.\\
	\textsuperscript{2}Pervasive Communication Research Center, Purple Mountain Laboratories, Nanjing 211111, China.\\
	
	\textsuperscript{*}Corresponding Authors
	\\
	{Email: \{yxyang, yilin\_ma\}@seu.edu.cn, changhengtai@pmlabs.com.cn, chxwang@seu.edu.cn}
	\vspace{-0.21 cm}	
}

\maketitle

\begin{abstract}
In this paper, a quasi-deterministic (Q-D) model for non-stationary underwater acoustic (UWA) channels is proposed. This model combines the BELLHOP deterministic model and geometry-based stochastic model (GBSM), which provides higher accuracy and flexibility. Different propagation components in shallow water are classified as D-rays, R-rays and F-rays in the proposed model, where D-rays are modeled by BELLHOP while both R-rays and F-rays are modeled by GBSM. Some important channel statistical properties, including time-frequency correlation function (TF-CF), Doppler power spectrum density (PSD), average Doppler shift, and RMS Doppler spread are derived and simulated. Finally, simulation results illustrate the correctness of the proposed model.  
\end{abstract}


\section{Introduction}
To achieve global coverage in 6G wireless communication systems, UWA communications is receiving increasing attention in recent years \cite{ref1,ref2,ref3}. Thus, as the first step to design UWA communication systems, accurate and general models for UWA channels are of great importance.

So far, there have been many studies on channel modeling \cite{ref5,ref6,ref7,ref8,ref9,ref10,ref11,ref12,ref13}. Those existing models are generally divided into three types, i.e., deterministic models, stochastic models, and Q-D models. For UWA channels, deterministic models like ray-tracing models \cite{ref5,ref6} and stochastic models like GBSMs \cite{ref7,ref8,ref9} are widely used. In \cite{ref5}, the effects of multipath and Doppler are studied under the application of ray-tracing method. In \cite{ref6}, BELLHOP tool was used to model four different frequency bands and two types of sea bottom for UWA channels. These models always have high accuracy but lack of flexibility. Compared with deterministic models, stochastic models, especially GBSMs, have higher flexibility. These models calculate distances and delays through geometry method, and can fit different scenarios by modifying the model parameters \cite{ref7,ref8,ref9}. In \cite{ref7,ref8,ref9}, three kinds of GBSMs considering different conditions for shallow water scenario were proposed. Although GBSM can fit different scenarios due to its high flexibility, its accuracy may not be suitable for some specific scenarios that require high accuracy and proper flexibility.
Q-D models are obtained by combining deterministic models and stochastic models, which have more flexibility than deterministic models and higher accuracy than stochastic models. In \cite{ref10}, a Q-D model for an urban environment was introduced. In \cite{ref11}, a Q-D model for millimeter-wave channels was proposed. In \cite{ref12}, a detailed tutorial on how to develop a Q-D model was provided. In \cite{ref13}, two kinds of method for simplifying the computational complexity were proposed.

Q-D model can achieve high accuracy and acceptable flexibility. However, to the best of the author’s knowledge, Q-D model is rarely used in underwater scenarios. To fill this gap, a Q-D model combining BELLHOP and GBSM for UWA channels is proposed in this paper.

The reminder of this paper is organized as follows. In Section II, a Q-D model combining BELLHOP and GBSM for UWA channels is proposed. In addition, the derivation of channel transfer function (CTF) is introduced. In Section III, the expressions of channel statistical properties are presented. Section IV presents the simulation results of the mentioned channel properties. The conclusions are finally drawn in Section V. 

\section{A Quasi-Deterministic Model Combining BELLHOP and GBSM}
\begin{figure}[htbp]
	\includegraphics[width=1.0\linewidth]{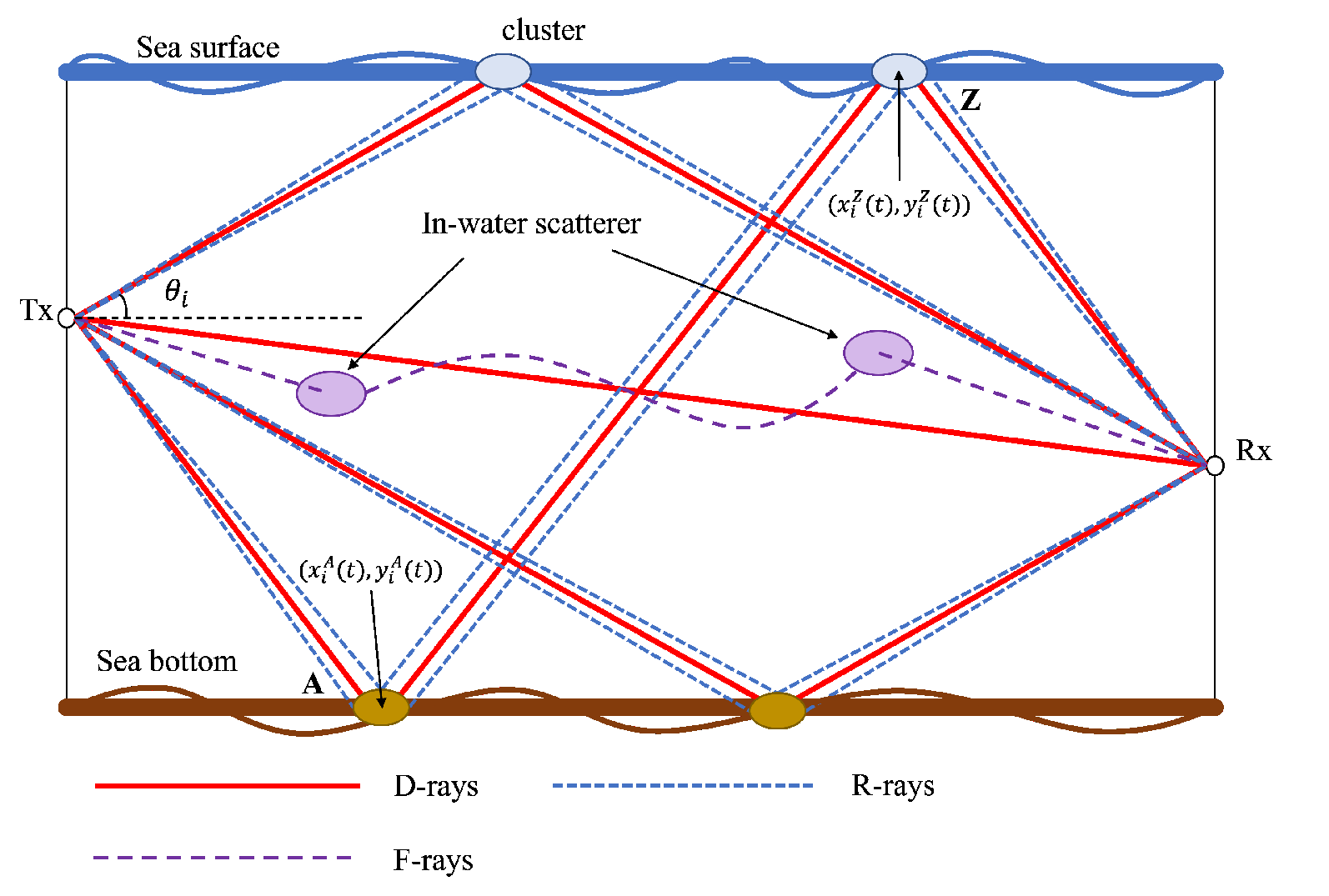}
	\caption{Different components of Q-D model}
	\label{fig1}
\end{figure}

The model is designed for UWA channels in shallow water, and both transmitter (Tx) and receiver (Rx) have no autonomous motion. In shallow water scenario, the speed of acoustic wave can be considered as constant and can propagate in a straight line \cite{ref17}. Fig. \ref{fig1} presents different components of Q-D model for UWA channels, including D-rays, R-rays and F-rays. D-rays represent reflection rays between the two boundaries and the LoS component, which can be generally considered stable in this scenario.  According to the experiment results, the number of collisions between rays and the sea surface and bottom, denoted by $N_\mathrm{S}$ and $N_\mathrm{B}$, usually satisfies $2 N_\mathrm{S}+2 N_\mathrm{B} \leq 8$ \cite{ref14,ref15,ref16}. Because of the roughness of boundaries, the rays are scattered and distributed around the reflection point, these diffuse scattering rays are modeled as R-rays. In addition, there are also multipath components caused by scatterers in water, which are modeled as F-rays. Different with R-rays, the clusters of F-rays are modeled as twin-cluster pairs, and the twin-clusters of F-rays will experience birth and death process over time.

For UWA channels, the channel impulse response (CIR) considering Doppler effect can be written as
\begin{equation}
	h(t, \tau)=\sum_{i=1}^N A_i {\delta}(t-\tau_i) e^{j 2 \pi a_i f_c t}
\end{equation}
where $A_i$ , $\tau_i$ and $a_i$ are channel gain, delay and Doppler factor of the $i$-th path, respectively. By applying the Fourier transform to the delay, the CTF can be obtained.
The total CTF of UWA channels can be represented by the superposition of CTFs of the D-rays, R-rays and F-rays components, and is given by
\begin{equation}
	H(t, f)=\sqrt{S_\mathrm{D}}H_\mathrm{D}(t,f)+\sqrt{S_\mathrm{R}}H_\mathrm{R}(t, f)+\sqrt{S_\mathrm{F}}H_\mathrm{F}(t, f)
\end{equation}
where $S_D$, $S_R$ and $S_F$ are power coefficients for D-rays, R-rays and F-rays, respectively.

\subsection{D-rays Modeling}

The CTF of D-rays can be written as
\begin{equation}
	H_\mathrm{D}(t,f)=\sum_{i=1}^N K_i A_i^\mathrm{D}(f) e^{-j 2 \pi(f_c+f) \tau_i^\mathrm{D}} e^{j 2 \pi a_i f_c t}
\end{equation}
where $N$ is the number of D-rays, $K_i$ represents the correction factor, which is similar to the Ricean factor, and is defined as $K_1 = \sqrt{K/K+1}, K_i = \sqrt{1/{(N-1)(K+1)}} (i \neq 1)$. The frequency-related channel gain is represented by $A_i^\mathrm{D}(f)$,  $f_c$ and $f$ represent the carrier frequency and signal frequency, respectively. The delay is represented by $\tau_i^\mathrm{D}$. All these parameters needed for modeling are obtained by BELLHOP simulation. Moreover, $a_i$ is the Doppler factor, and will be introduced at the end of this section, and $t$ represents the time instant. In addition, in order to facilitate the simulation of the CTFs of R-rays and F-rays, parameters including the angle of the $i$-th D-ray in the azimuth direction, represented by $\theta_i$, the number of the $i$-th D-ray contacting with the surface (bottom), represented by $s_i$($b_i$), and coordinates of the first reflection point $\rm{A}$ on the $i$-th path relative to Tx as well as the last reflection point $\rm{Z}$ on the $i$-th path relative to Rx, i.e., $\left(x_i^\mathrm{A}, y_i^\mathrm{A}\right)$ and $\left(x_i^\mathrm{Z}, y_i^\mathrm{Z}\right)$ are also obtained through BELLHOP simulation. 

\subsection{R-rays Modeling}

For R-rays, the CTF is presented as
\begin{equation}
	H_\mathrm{R}(t, f)=\sum_{i=2}^N \sum_{j=1}^M A_{i, j}^\mathrm{R}(t, f) e^{-j 2 \pi\left(f_c+f\right) \tau_{i, j}^\mathrm{R}(t)}e^{j 2 \pi a_i f_c t}
\end{equation}
where $M$ is the number of scattering rays. The index of path starts at 2, since the $1$th path represents the LoS path.
To calculate the delay of each scattering ray, the total distance of the $j$-th scattering ray around the $i$-th D-ray $d_{i, j}^\mathrm{R}(t)$ is classified into three categories, i.e., $d_{i, j}^\mathrm{R}(t)=d_{i, j}^\mathrm{R,A}(t)+d_{i, j}^\mathrm{R,Z}(t)+d_{i, j}^{\Delta}(t)$, where $d_{i, j}^\mathrm{R,A}(t)$ and  $d_{i, j}^\mathrm{R,Z}(t)$ represent the first bounce distance between Tx and the $j$-th scattering point around the reflection point $\rm{A}$ of the $i$-th D-ray and the last bounce between Rx and the $j$-th scattering point around the reflection point $\rm{Z}$ of the $i$-th D-ray, respectively. The distance between the first and last scatterers is $d_{i, j}^{\Delta}(t)=\left(D_i-d_{i, j}^\mathrm{R,A}(t)-d_{i, j}^\mathrm{R,Z}(t)\right) \cdot e^{\Delta d}$, where $D_i$ is the total distance of the $i$-th D-rays, and is obtained by $D_i(t)=v \cdot \tau_i(t)$, and $\Delta d$ is a random variable satisfying zero-mean Gaussian distribution, i.e., $\Delta d \sim N\left(0, \sigma_d^2\right)$. For the single-bounce scattering paths ($s_i + b_i = 1$), $d_{i, j}^\mathrm{R}(t)=d_{i, j}^\mathrm{R,A}(t)+d_{i, j}^\mathrm{R,Z}(t)$.

To calculate the distance of the scattering paths, the positions of scatterers around the reflection point can be regarded as following Gaussian distribution \cite{ref18}, i.e.,
 \begin{equation}
 	p\left(x_{i, j}^\mathrm{A(Z)}, y_{i, j}^\mathrm{A(Z)}\right)=\frac{\exp \left[-\frac{\left(x_{i, j}^\mathrm{A(Z)}-x_i^\mathrm{A(Z)}\right)^2}{2 \sigma_x^2}-\frac{\left(y_{i, j}^\mathrm{A(Z)}-y_i^\mathrm{A(Z)}\right)^2}{2 \sigma_y^2}\right]}{2 \pi \sigma_x \sigma_y}
 \end{equation}
where $\sigma_x$ and $\sigma_y$ are standard deviations of the Gaussian distributions. Here, $x_i^\mathrm{A(Z)}$ and $y_i^\mathrm{A(Z)}$ represent $x$ and $y$ coordinate of the first (last) reflection point. The relative coordinates centered at the reflection point A(Z) is represented by $\left(x_{i,j}^\mathrm{A(Z)}(t), y_{i,j}^\mathrm{A(Z)}(t)\right)$. Then the distance of the first-bounce (last-bounce) can be calculate by
\begin{equation}
	d_{i, j}^\mathrm{A(Z)}(t)=\sqrt{\left(x_i^\mathrm{A(Z)}+x_{i, j}^\mathrm{A(Z)}(t)\right)^2+\left(y_i^\mathrm{A(Z)}+y_{i, j}^\mathrm{A(Z)}(t)\right)^2}.
\end{equation}
 Finally, the delay of each scattering ray can be obtained as 
\begin{equation}
	\tau_{i, j}^\mathrm{R}(t) = d_{i,j}^\mathrm{R}(t)/v
\end{equation}
where $v$ is the speed of acoustic wave in shallow water.

The channel gain can be divided into 4 parts, i.e.,
\begin{equation}
	A_{i, j}^\mathrm{R}(t, f)= L_\mathrm{S}\left(d_{i, j}^\mathrm{R}(t)\right) L_\mathrm{A}\left(d_{i, j}^\mathrm{R}(t), f\right) L_\mathrm{B}\left(\varphi_i(t)\right)^{b_i} {L_R}^{s_i} .
\end{equation}
Here,  $L_\mathrm{S}(d)=1 / \sqrt{d^\beta}$ represents the spreading loss, where $d$ is the propagation distance and $\beta$ is spreading factor. The absorption loss is given by \cite{ref19}
\begin{equation}
L_\mathrm{A}(d, f)=10^{-\frac{d \cdot \alpha(f)}{20000}} 
\end{equation}
where $\alpha(f)$ is the absorption factor, and can be given by Thorp model, i.e.,\cite{ref20}
\begin{equation}
 \alpha(f)=\frac{0.11 f^2}{1+f^2}+\frac{44 f^2}{4100+f^2}+2.75 \times 10^{-4} f^2+0.003.
\end{equation}
The bottom reflection loss $L_\mathrm{B}(\varphi)$ is given by \cite{ref20}
\begin{equation}
 L_\mathrm{B}(\varphi)=\left|\frac{\left(\rho_b / \rho_w\right) \cos (\varphi)-\sqrt{\left(c_w / c_b\right)^2-\sin ^2(\varphi)}}{\left(\rho_b / \rho_w\right) \cos (\varphi)+\sqrt{\left(c_w / c_b\right)^2-\sin ^2(\varphi)}}\right|.
\end{equation}
Here, $\rho_b$ and $\rho_w$ represent density of sea bottom and water, respectively. Similarly, the speed of acoustic wave in sea bottom and water is represented by $c_b$ and $c_w$, respectively.  The characteristic angle of incidence $\varphi_i$ can be calculated by $\varphi_i(t)=\frac{\pi}{2}-\theta_i(t)$. Since the sea surface reflection is regarded as specular reflection, it is represented by $L_R = -1$.

\subsection{F-rays Modeling}

For F-rays, the CTF can is presented as
\begin{equation}
	H_\mathrm{F}(t, f)=\sum_{i=1}^{N_\mathrm{F}} \sum_{j=1}^{M_\mathrm{F}} A_{i, j}^\mathrm{F}(t, f) e^{-j 2 \pi\left(f_c+f\right) \tau_{i, j}^\mathrm{F}(t)}e^{j 2 \pi a_i^\mathrm{F} f_c t}
\end{equation}
where $N_\mathrm{F}$ is the number of in-water clusters, $M_\mathrm{F}$ represents the number of scattering rays. Moreover, $a_i^\mathrm{F}$ is the Doppler factor for multipaths in water.

To get the delay of each scattering ray, the positions of clusters in water need to be calculated. Here, the angle $\phi_\mathrm{F}^\mathrm{A(Z)}$ and distance $d_\mathrm{F}^\mathrm{A(Z)}$ of the cluster relative to the Tx (Rx) are derived through birth-death model. The description of birth-death process is mainly based on the survival probability. Then for the Tx (Rx) side, the probability of a cluster surviving over the time difference $\Delta t$ is \cite{ref18}
\begin{equation}
	P_s^{T(R)}(\Delta t)=\exp \left[-\lambda_R\left(\frac{\Delta t \cdot v_{t(r)}}{D_c^S}\right)\right].
\end{equation} 
Here, $\lambda_R$ represents the recombination rate of in-water clusters, $D_c^S$ represents the scenario-related correlation parameter. In this paper, the Tx and Rx have no autonomous motion speed, so $v_{t(r)}$ will take a small value, which indicates the drifting speed of the Tx and Rx. Then, the combined probability of the survival of clusters can be given by \cite{ref18}
\begin{equation}
	P_s(\Delta t)=P_s^T(\Delta t) \cdot P_s^R(\Delta t)
\end{equation} 
and the average number of new clusters is \cite{ref18}
\begin{equation}
	\mathbb{E}\left(N_{\text {new }}\right)=\frac{\lambda_G}{\lambda_R}\left(1-P_s(\Delta t)\right).
\end{equation}

The total distance of each ray is
\begin{equation}
  d_{i, j}^\mathrm{F}(t)=d_{i, j}^\mathrm{F, A}(t)+d_{i, j}^\mathrm{F, Z}(t)+\tilde{\tau}_{i, j} \cdot c.
\end{equation} 
Here, $d_{i, j}^\mathrm{F, A}(t)$ and $d_{i, j}^\mathrm{F, Z}(t)$ represent the distance of Tx to the first in-water scatterer and Rx to the last in-water scatterer, respectively. The delay between two clusters can be written as $\tilde{\tau}_{i, j}=\frac{\tilde{d_{i, j}}}{c}+\tau_{c, l i n k}$, where $\tilde{d_{i, j}}$ is the distance between the two clusters, and $\tau_{c, l i n k}$ is a random variable that follows exponential distribution \cite{ref18}. Then the delay is calculated by $\tau_{i, j}^\mathrm{F}(t)=\frac{d_{i, j}^\mathrm{F}(t)}{c}$, and the channel gain is given by 
\begin{equation}
 A_{i, j}^\mathrm{F}(t, f)=L_\mathrm{S}\left(d_{i, j}^\mathrm{F}(t)\right) L_\mathrm{A}\left(d_{i, j}^\mathrm{F}(t), f\right).
\end{equation} 

\subsection{Doppler Effect}

The relative motion of Tx, Rx, and scatters lead to Doppler effect. In this paper, two types of motion leading to the Doppler effect are considered, i.e., unintentional Tx/Rx motion, which is regarded as drifting and gives rise to the Doppler factor $a_{d p}$, and the motion of sea surface, which gives rise to the Doppler factor $a_{sp}$ \cite{ref4}.
Assuming that the time-varying drifting speed of Tx (Rx) is $v_t$ ($v_r$), the angle between the drifting direction and the horizontal direction of Tx (Rx) is $\theta_t$ ($\theta_r$), and the motion direction of the $i$-th D-ray is $\theta_i$. Then the relative speed and the corresponding Doppler factor can be written as \cite{ref4}
\begin{equation}
	v_d=\left[v_t \cos \left(\theta_i-\theta_t\right)-v_r \cos \left(\theta_i+\theta_r\right)\right]
\end{equation}
\begin{equation}
	a_{d, i}=\frac{v_d}{c}.
\end{equation}

To calculate $a_{sp}$, the waves are modeled to move in a sinusoidal pattern with amplitude $A_w$ and frequency $f_w$. Then for the $i$-th path, the vertical speed after the $j$-th sea surface reflection can be expressed as \cite{ref4}
\begin{equation}
	v^{\prime}=v_w \sin \left(\psi_{i, j}+2 \pi f_w t\right)
\end{equation}
where $v_w=2 \pi f_w A_w$, $\psi_{i,j}$ is uniformly distributed over $[-\pi, \pi]$.  Then for the $i$-th path, the total vertical speed after $s_i$ sea surface reflections is \cite{ref4} 
\begin{equation}
	v_{\perp}=2 v_w \sin \theta_i \sum_{j=1}^{S_i} \sin \left(\psi_{i, j}+2 \pi f_w t\right).
\end{equation}
Then, the Doppler factor for sea surface motion is written as 
\begin{equation}
	a_{s p, i}=\frac{v_{\perp}}{c}.
\end{equation}

For D-rays and R-rays, the Doppler components include drifting and surface motion, while for F-rays, the only component is drifting, i.e.,
\begin{subequations}
	 \begin{equation}
		a_i=a_{d, i}+a_{s p, i}
	\end{equation}
	\begin{equation}
		a_i^\mathrm{F}=a_{d, i}.
	\end{equation}
\end{subequations}

\section{Statistical Properties}

\subsection{TF-CF}
For the proposed model, the definition of TF-CF is
\begin{equation}
	R_H(t, f ; \Delta t, \Delta f)=\mathrm{E}\left\{H(t, f) H^*(t+\Delta t, f+\Delta f)\right\}
\end{equation}
Here, $\mathrm{E}\{\cdot\}$ represents statistical average and $(\cdot)^*$ represents complex conjugation. By substituting (3), (4), and (12) into (24), the TF-CF is obtained by
\begin{equation}
	\begin{aligned}
	    R_H(t, f ; \Delta t, \Delta f)=S_\mathrm{D}\sum_{i=1}^N R_{H, i}^\mathrm{D}(t, f ; \Delta t, \Delta f)
	    \\+S_\mathrm{R} \sum_{i=2}^N  R_{H, i }^\mathrm{R}(t, f ; \Delta t, \Delta f)
	    \\+S_\mathrm{F} \sum_{i=1}^{N_F} R_{H, i}^\mathrm{F}(t, f ; \Delta t, \Delta f).
	\end{aligned}
\end{equation}
\begin{figure*}[ht]
	\begin{subequations}
		\begin{equation}
			R_{H,i}^\mathrm{D}\left( {t,f;\mathrm{\Delta}t,\mathrm{\Delta}f} \right) = E\left\{\frac{1}{N} {{G_{i}^\mathrm{D}e}^{ j2\pi f_c a_{i}\mathrm{\Delta}t}} \right\}
		\end{equation}
		\begin{equation}	
			R_{H,i}^\mathrm{R}\left( {t,f;\mathrm{\Delta}t,\mathrm{\Delta}f} \right) = E\left\{ { {\left( {\sum\limits_{j = 1}^{M}\frac{1}{N}{G_{i,j}^\mathrm{R}e^{- j2\pi (f_c+f){\lbrack{\tau_{i,j}^\mathrm{R}{(t)} - \tau_{i,j}^\mathrm{R}{({t + \mathrm{\Delta}t})}}\rbrack} - j2\pi\mathrm{\Delta}f\tau_{i,j}^\mathrm{R}{({t + \mathrm{\Delta}t})}}}} \right)e}^{ j2\pi f_ca_{i}\mathrm{\Delta}t}} \right\}
		\end{equation}
		\begin{equation}
			R_{H,i}^\mathrm{F}\left( {t,f;\mathrm{\Delta}t,\mathrm{\Delta}f} \right) = E\left\{ { {\left( {\sum\limits_{j = 1}^{M_\mathrm{F}}\frac{1}{N_\mathrm{F}}{G_{i,j}^\mathrm{F}e^{- j2\pi (f_c+f){\lbrack{\tau_{i,j}^\mathrm{F}{(t)} - \tau_{i,j}^\mathrm{F}{({t + \mathrm{\Delta}t})}}\rbrack} - j2\pi\mathrm{\Delta}f\tau_{i,j}^\mathrm{F}{({t + \mathrm{\Delta}t})}}}} \right)e}^{ j2\pi f_ca_{i}^\mathrm{F}\mathrm{\Delta}t}} \right\}
		\end{equation}
	\end{subequations}
\end{figure*}

The expression of correlation functions of three kinds of rays are given at the top of the next page, where $G_{i,j}^\mathrm{X}$ represents the normalized correlation function of amplitude, in which $\mathrm{X}=\left\{\mathrm{D,R,F}\right\}$. Then the temporal autocorrelation function (ACF) can be obtained by setting $\Delta f$ to 0.

\subsection{Coherence Time}
Coherence time represents the maximum time that the channel can be regarded as constant. It can be defined as \cite{ref21}
\begin{equation}
	T(t,f) = \mathrm{min}\{\Delta t>0: R_H(t,f;\Delta t) = c_\mathrm{thresh\_T}\}.
\end{equation}
Here, $c_\mathrm{thresh\_T}$ is a threshold in [0,1].

\subsection{Doppler PSD}
Doppler PSD $\phi_H(t,\nu)$ illustrates the relationship between the average power and Doppler frequency $\nu$, which can be obtained by 
\begin{equation}
	\phi_H(\nu)=\int_{-\infty}^{\infty} R_H(t,f;\Delta t) e^{-j 2 \pi \Delta t \nu} d \Delta t.
\end{equation}

\subsection{Average Doppler shift and RMS Doppler spread}
The expression of average Doppler shift, represented by $\mu_\nu$ and RMS Doppler spread, represented by $\sigma_\nu$ is given by
\begin{subequations}
	\begin{equation}
		\mu_\nu=\frac{\int_{-\infty}^{\infty} \nu \phi_H(t,\nu) d \nu}{\int_{-\infty}^{\infty} \phi_H(t,\nu) d \nu} 
	\end{equation}
	\begin{equation}
		\sigma_\nu=\sqrt{\frac{\int_{-\infty}^{\infty}\left(\nu-\mu_\nu\right)^2 \phi_H(\nu) d \nu}{\int_{-\infty}^{\infty} \phi_H(\nu) d \nu}}.
	\end{equation}
\end{subequations}
\section{Results and Analysis}
 In this simulation, the water depth is 100 m, propagation distance is 1500 m, and the depths of Tx and Rx are $h_t = 40$ m and $h_r = 60 $ m.  The speed of acoustic wave in water is 1500 m/s, and in the bottom it is 1600 m/s. The velocities of Tx and Rx ($v_t$ and $v_r$) are modeled as time-varying sinusoidal functions with amplitudes of 0.1 m and 0.02 m. The motion directions of Tx and Rx ($\theta_t$ and $\theta_r$) satisfy the random distribution on $[0,2\pi]$.
\begin{figure}[htbp]
	\centering
	\includegraphics[width=0.8\linewidth]{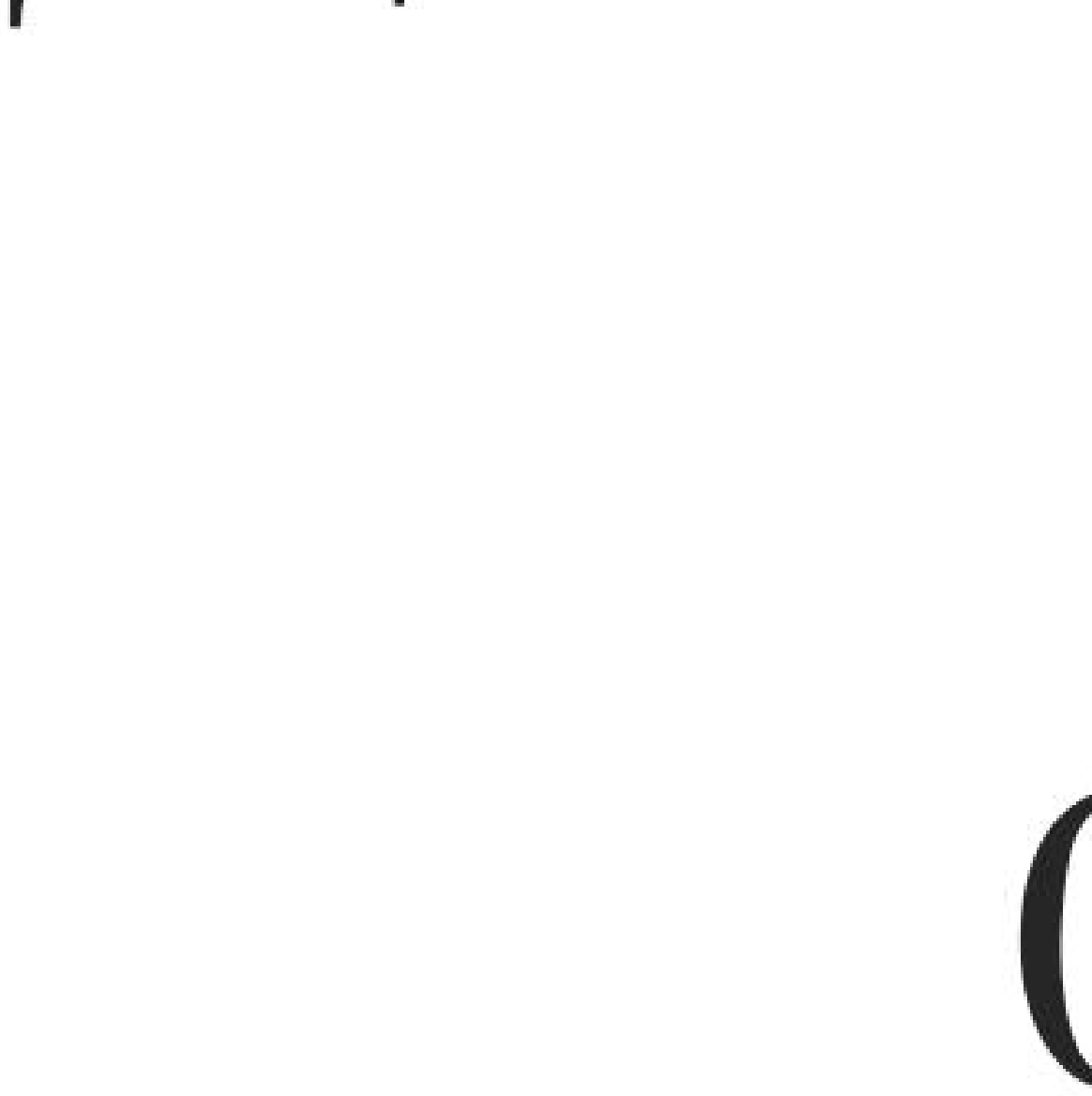}
	\caption{Temporal ACFs at different times and carrier frequencies ($ K = 1.6, S_\mathrm{D} = 0.4, S_\mathrm{R} = 0.4, S_\mathrm{F} = 0.2$).}
	\label{fig2}
\end{figure}

Fig. \ref{fig2} gives temporal ACFs at two time instants and three carrier frequencies. The temporal ACFs decrease rapidly with the increase of time difference in all the conditions. The difference of ACFs at different instants reflects the non-stationarity of UWA channels in the time domain. In addition, with the increase of the carrier frequency, the temporal ACFs decrease faster, which illustrates that the magnitude of coherence time is associated with carrier frequency, i.e., the higher carrier frequency contributes to shorter coherence time.

\begin{figure}[htbp]
	\centering
	\includegraphics[width=0.8\linewidth]{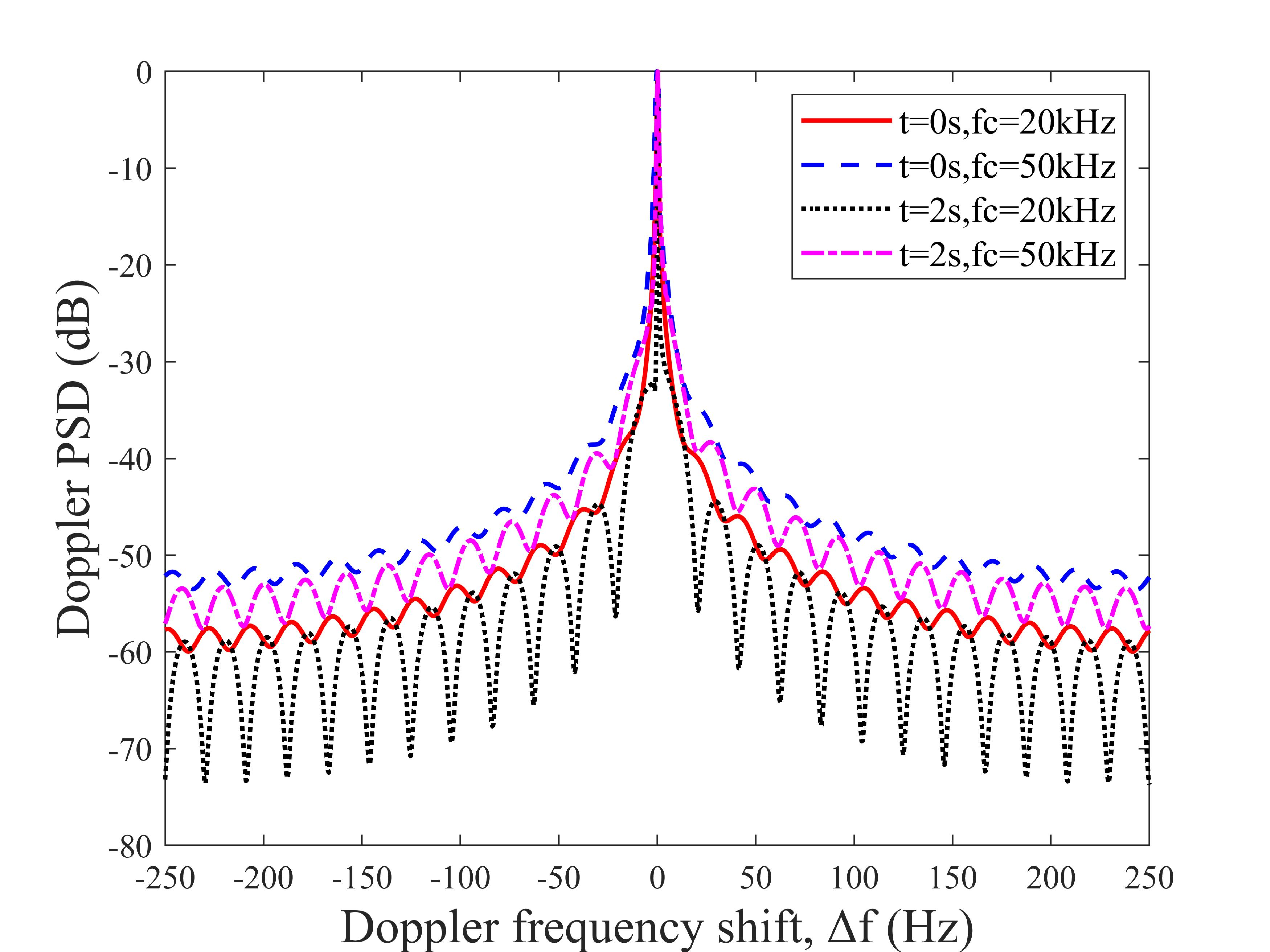}
	\caption{Doppler PSDs at different times and carrier frequencies ($ K = 1.6, S_\mathrm{D} = 0.4, S_\mathrm{R} = 0.4, S_\mathrm{F} = 0.2$).}
	\label{fig3}
\end{figure}

\begin{table}[htbp]
	\centering
	\caption{Average Doppler shift and RMS Doppler spread ($ K = 1.6, S_\mathrm{D} = 0.4, S_\mathrm{R} = 0.4, S_\mathrm{F} = 0.2$).}
	\renewcommand{\arraystretch}{1.5}
	\large
	\resizebox{\linewidth}{!}{
		\begin{tabular}{|c|c|c|c|c|} \hline		
			\diagbox{Results}{Conditions} &0s 20kHz &0s 50kHz &2s 20kHz &2s 50kHz	 \\ \hline
			Average Doppler shift (Hz)	& 0.0369	&0.1161	&0.0074  &0.0384 \\ \hline
			RMS Doppler spread (Hz)	& 150.9	&151.7 &150.4  &151.3 \\ \hline
		\end{tabular}
	}	
\end{table}

Fig. \ref{fig3} gives the simulation results of Doppler PSD. Doppler PSD changes over time, indicating that the random relative motion at different time instant of Tx and Rx lead to different Doppler spreads. It can also be found in the simulation results that the higher carrier frequency contributes to the larger Doppler spread. The results of average Doppler shift and RMS Doppler spread are given in Table I. Because of the fact that both Tx and Rx have no autonomous motion speed, the main contribution to the Doppler is the drifting, resulting in a small average Doppler shift.

\section{Conclusion}

	In this paper, a new Q-D model combining BELLHOP and GBSM for non-stationary UWA channels in shallow water has been proposed. A new classification of rays in underwater scenario has been presented, i.e., D-rays , R-rays and F-rays. In this model, CTFs of three kinds of rays have been derived by BELLHOP and GBSM, respectively. Then, the derivation and simulation of some important channel properties including temporal ACF, Doppler PSD, average Doppler shift and RMS Doppler spread have been introduced. The fact that temporal ACF changes over time illustrates the time non-stationarity in shallow water scenario of UWA channels, and as the carrier frequency increases, the coherence time decreases. Moreover, Doppler shift and RMS Doppler spread are affected by random motions of Tx and Rx at different times, and the latter get larger with the increase of carrier frequency.

\section*{Acknowledgement}
This work was supported by the National Natural Science Foundation of China (NSFC) under Grant 61960206006 and 62301365, the Fundamental Research Funds for the Central Universities under Grant 2242022k60006, the Key Technologies R\&D Program of Jiangsu (Prospective and Key Technologies for Industry) under Grants BE2022067, BE2022067-1 and BE2022067-3, and the EU H2020 RISE TESTBED2 project under Grant 872172.

\end{document}